\documentclass{ws-fnl2}
\usepackage{cite}
\usepackage{amsmath}
\usepackage[subnum]{cases}
\usepackage{multirow}
\begin{document}

\markboth{E. Gonzalez \& L.B. Kish}{Key Exchange Trust Evaluation in Peer-to-Peer Sensor Networks with Unconditionally Secure Key Exchange}

\catchline{}{}{}{}{}

\title{KEY EXCHANGE TRUST EVALUATION IN PEER-TO-PEER SENSOR NETWORKS WITH UNCONDITIONALLY SECURE KEY EXCHANGE}

\author{\footnotesize ELIAS GONZALEZ, LASZLO B. KISH}

\address{Electrical and Computer Engineering, Texas A\&M University, 3128 TAMU\\
College Station, TX 77843-3128, United States of America\\
eliasg23@tamu.edu, laszlo.kish@ece.tamu.edu}

\maketitle

\begin{history}
\received{(received date)}
\revised{(revised date)}
\end{history}

\begin{abstract}
As the utilization of sensor networks continue to increase, the importance of security becomes more profound. Many industries depend on sensor networks for critical tasks, and a malicious entity can potentially cause catastrophic damage. We propose a new key exchange trust evaluation for peer-to-peer sensor networks, where part of the network has unconditionally secure key exchange. For a given sensor, the higher the portion of channels with unconditionally secure key exchange the higher the trust value. We give a brief introduction to unconditionally secured key exchange concepts and mention current trust measures in sensor networks. We demonstrate the new key exchange trust measure on a hypothetical sensor network using both wired and wireless communication channels.
\end{abstract}

\section{Introduction}

\subsection{Sensor networks}

Sensor networks consist of sensors that measure and provide information in remote or spatially distributed areas \cite{intro1}. With the advancement of miniaturization and wireless technologies, the ubiquity of sensor networks is becoming more prevalent. The benefits of having smaller dies in semiconductors include; physically smaller devices, increase ratio of computing power per energy, better battery life, etc. A few examples that utilize sensor networks include military, health care, environment monitoring, agriculture, etc.

Sensors are often required to be autonomous, decentralized, and in remote areas. Such requirements place limitations on sensors and sensor networks, including low power, limited memory and data storage, physical size, limited communication bandwidth, cost, privacy, and security \cite{intro2, intro3, intro4}. There are proposed solutions for some of these limitations such as energy harvesting, low-power processors, smaller memory footprint, etc. However, security is a pressing issue since sensors face unique challenges. Without proper security the entire sensor network can be compromised and sabotaged.

\subsection{Security concerns}

Limited computing power in sensors restrict them from utilizing large complex encryption algorithms, also with limited memory and data storage the secure key cannot be too large. Another security issue facing sensors is that the installation of optical fiber or wire connections is often not economical. Thus they are often accessible only by wireless communication, which is restricted to work with conditionally secure key exchange, and make them vulnerable to packet capture, sniffing, and injection \cite{sec1, sec2, sec3, sec4, sec5, sec6, sec7, sec8}. In an attempt to mitigate some of these security issues, there have been several proposals to secure sensor networks, which include defenses against specific attacks and more efficient protocols \cite{sec1, sec2, sec3, sec4, sec5, sec6, sec7, sec8}.

Sensor networks require data confidentiality, data integrity, data freshness, availability, self-organization, time synchronization, authentication, secure broadcasting and multicasting, and sensor privacy. Attacks on sensor networks include Denial of Service (DoS) attacks, Sybil attacks, traffic analysis attacks, information flooding attacks, and node replication attacks \cite{sec1, sec2, sec3, sec4, sec5, sec6, sec7, sec8}. Defensive measures against some of these attacks are key establishment, key encryption, policy-based approaches, intrusion detection, and trust management. There have been several approaches for managing trust in sensor networks, the approach to trust management is based on the sensor network's trust mechanism.

\subsection{Trust mechanisms}

Trust theory has different applications and perspectives, and the concept of trust has been associated with past behaviors and/or reputation from trusted peers \cite{trust1, trust2, trust3, trust4, trust5, trust6}. The notion of trust has been specified by trust definitions, trust characteristics, and trust values \cite{trust7}. Trust values have been measured by several different methodologies such as; Bayesian models \cite{bayesian}, Beta distribution systems \cite{beta}, subjective logic models \cite{subjective}, entropy models \cite{entropy}, fuzzy models \cite{fuzzy}, and game theory models \cite{gametheory}. However, these trust value models are not able to distinguish between conditional and unconditionally secure key exchanges, thus these models need to be expanded for related applications.

Rather than expanding former models we propose a new key exchange trust evaluation model, which takes into account the type of key exchange (conditional/unconditional) between two sensors.

\subsection{Unconditionally secure key exchange}

In software-based key distribution (exchange) protocols the security is only computationally-conditional, meaning that the eavesdropper has all the communicated information, and with enough computing resources or time the key can be fully extracted. The advantage of software-based key distributions is that they are relatively cheap, easy to install and run, and the key can be exchanged wirelessly. 

Unconditionally secure key exchanges are key distribution methods that are information theoretically secure \cite{infotheosec}, which means that the information is not in the communicated signal, see next paragraph. Thus even with infinite computing resources the eavesdropper cannot extract the key. However, physical (hardware-based) key exchanges are the only schemes that can provide unconditionally secure key exchange. Hardware-based key exchanges are more expensive than software-based schemes, moreover, wireless key exchange is not possible (except quantum key distribution with single photons, which require complete darkness.)

So far there are two physical key distribution classes that offer unconditionally secure key exchange: Quantum Key Distribution(QKD) \cite{bb84} and the Kirchhoff-Law-Johnson-Noise(KLJN) scheme \cite{kljn1, lk1, lk2, lk3, kljn_its, kljn_noise_properties, kljn_noise_properties2, barry1, barry2, c186, c187a, c174, c171, c169, c147, c141, c133, c128, c118, c111, c113, c+6, barry3}.

In QKD principle, the bits are exchanged via single photon communications and the physical law which provides unconditional secure key exchange is the quantum no-cloning theorem \cite{bb84}. Recently, the fundamental security proofs for QKD have been debated \cite{yuen1, yuen2, hirota}. QKD has also had issues with the non-ideality of practical building elements, which have lead to the cracking of existing communicators, including commercial devices \cite{crack1, crack2, crack3, crack4, crack5, crack6, crack7, crack8, crack9, crack10}. Although, these practical non-ideality problems can be patched there is no security until the patch is known and applied. Other concerns with QKD systems are the bulky physical size, it is relatively expensive, requires large power consumption, its sensitivity to vibrations, and the required ``dark optical fiber''. These characteristics of QKD make it almost impossible to integrate into a sensor. 

In the KLJN scheme, the key bit is exchanged via a wire channel and utilizes statistical physics \cite{kljn1}. The actual physical laws of providing security are the second law of thermodynamics and the properties of Gaussian fluctuations. Relative to QKD, KLJN can be integrated on a microchip thus it does not have issues with physical size, energy required, sensitivity to vibrations, etc. KLJN can be implemented into a sensor, but will require a wire to connect every sensor that intends to acquire a unconditionally secure key exchange.

\begin{figure}[ht]
    \label{figone}
  \centering
    \includegraphics[width=0.65 \textwidth]{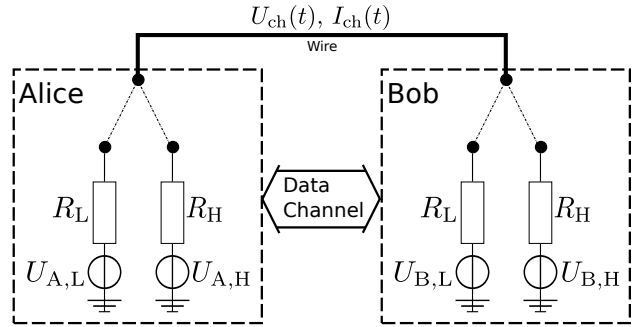}
    \vspace*{8pt}
    \caption{An illustration of the core KLJN system. Alice and Bob each have a communicator which have noise generators, a ``Low'' resistor $R_\mathrm{L}$ (representing the Low bit value), and a ``High'' resistor $R_\mathrm{H}$ (representing the High bit value.) The noise voltages are enhanced by generators emulating Johnson noise $U_\mathrm{A,L}$ or $U_\mathrm{A,H}$ for Alice; and $U_\mathrm{B,L}$ or $U_\mathrm{B,H}$ for Bob, at very high temperature. Once the communicators select a resistor they measure the mean-squared voltage amplitude $<U^{2}_\mathrm{ch}(t)>$ and/or the current amplitude $<I^{2}_\mathrm{ch}(t)>$. There is a wire for the key exchange, and there is a channel for data exchange. Against active attacks and attacks exploiting component non-idealities, an authenticated public data channel is used to measure and compare bits \cite{lk2, c111}.}
\end{figure}

An illustration of the KLJN setup is in Figure~\ref{figone}. In this figure Alice and Bob have two identical resistor pairs which are $R_\mathrm{L}$ for the Low resistor and $R_\mathrm{H}$ for the High resistor. Each resistor has noise voltages that are enhanced by Johnson noise, $U_\mathrm{A,L}$ for Alice's Low resistor, $U_\mathrm{A,H}$ for Alice's High resistor, $U_\mathrm{B,L}$ for Bob's Low resistor, and $U_\mathrm{B,H}$ for Bob's High resistor. During the key bit exchange period the first step is for Alice and Bob to select either $R_\mathrm{L}$ or $R_\mathrm{H}$. The selection of $R_\mathrm{L}$ and $R_\mathrm{H}$ is random and both are equally likely to be selected. Since the selection of $R_\mathrm{L}$ and $R_\mathrm{H}$ is random neither Alice or Bob know which resistor will be selected. Once Alice and Bob select their respective resistor they measure the voltage and/or current in the wire. The channel voltage can be modeled by $<U^{2}_\mathrm{ch}(t)>=4\mathrm{k}T_\mathrm{eff}B_\mathrm{KLJN}$ and the channel current can be modeled by $<I^{2}_\mathrm{ch}(t)>=4kT_\mathrm{eff}B_\mathrm{KLJN}/R_\mathrm{loop}$ with k being Boltzmann's constant, $T_\mathrm{eff}$ measuring the effective temperature, $R_\mathrm{loop}$ being the loop resistance, and $B_\mathrm{KLJN}$ being the KLJN bandwidth \cite{kljn1}. From $<U^{2}_\mathrm{ch}(t)>$ or $<I^{2}_\mathrm{ch}(t)>$ Alice and Bob know which resistor the other end selected, and they already know which resistor they selected. If the voltage noise level is high then they both selected high resistors, and if the voltage noise level is low then they both selected low resistors, in these outcomes the key bit is discarded and the next period begins. If an intermediate voltage noise level or current noise level is measured then a secure key bit is generated, stored, and the next period begins. This process continues until the desired number of key bits are generated.

The core system in Figure~\ref{figone} is secure against passive (non-invasive) attacks in the idealized case. However in \cite{lk2}, when Eve is tampering with or changing the system via an active/invasive intervention such as launching a MITM (man-in-the-middle) attack \cite{c111}, the core system is not enough to guarantee security. Similarly, non-idealities which represent deviations from the original scheme, cause security leak \cite{lk2}. For defending the system against these kind of attacks, the instantaneous voltage and current amplitudes are measured by Alice and Bob, and these quantities are communicated and compared via a public authenticated data channel. Alice and Bob have a full and deterministic model of the system, because it is a classical physical system, therefore incessant measurement of the current and voltage is allowed. Based on their comparison and preconditions, Alice and Bob decide to keep or discard the bit having compromised security \cite{c111}. The authentication uses only log$_2(M)$ secure bits of the exchanged bits, where $M$ is the number of bits carrying the current and voltage data in the public channel. In practical applications this channel can be wireless or wired.

Utilizing KLJN in sensor networks could significantly increase the security level in sensor networks due to its unconditionally secure key exchange.

\subsection{Motivation for a key exchange trust evaluation}

Current trust measures for sensor networks do not utilize unconditionally secure key exchange. Trust is a belief that may change over time, and is usually based on past behaviors and/or reputation from a community. Many sensor networks measure trust based on past behaviors and/or reputation \cite{trust1, trust2, trust3, trust4, trust5, trust6}, but there has not been a trust measurement that considers the class (conditionally/unconditionally secure) of the key exchange utilized in their measurement of trust. We propose a new key exchange trust system that considers the class of the key exchange. The system utilizes the \textbf{G}eometric series to evaluate the key exchange trust, thus we call it the G key exchange trust function.

\section{Outline of Combined Wired and Wireless Sensor Networks}

In this paper we consider peer-to-peer networks only. In such a network it will be impractical to have direct wired connections from every sensor to every other sensor, thus we propose to use both wired and wireless communication channels, and form a wired-wireless hybrid network. The wired sensors can be utilized in areas where other sensors are in close proximity. Each sensor can then be ranked based on its key exchange and the number of key exchanges with trusted peers. We therefore propose the G key exchange trust measure system.

\subsection{Network}

The wired-wireless network will require sensors to have at least two communication devices, one for wireless and one or more for wired. A cable will also be required and can have either one or two wires inside. One wire will be for the key exchange, and the other optional wire can be utilized as a data communication channel. 

\begin{figure}[ht]
    \label{fig:network}
\centering
\includegraphics[width=0.75 \textwidth]{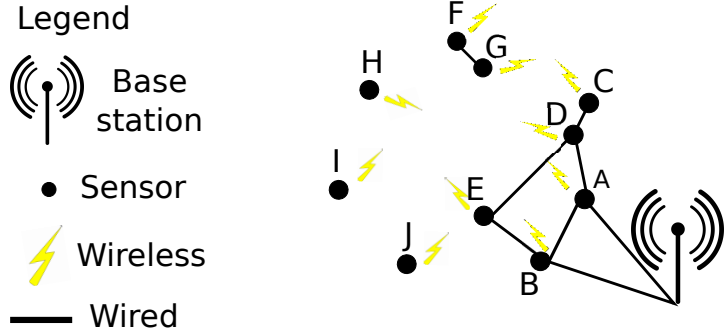}
\vspace*{8pt}
\caption{An illustration of a wired-wireless hybrid sensor network. In this example there are ten sensors with only select sensors utilizing wired communication channels and all sensors utilizing wireless communication channels.}
\end{figure}

Figure~\ref{fig:network} is an illustration and example of the proposed wired-wireless hybrid sensor network with ten sensors. In this example sensors A through G utilize both wired and wireless communication channels, and sensors H through J utilize only its wireless communication channel. Sensors A and B have a direct connection with the base station, thus they can have an unconditionally secure key exchange with the operator. Note how sensor E has two wired connections, this sensor will require two KLJN communicators. Sensors C, F, and G have only one wired connection and will require one KLJN communicator. Sensors A, B, and D have three wired connections, and will require three KLJN communicators. Sensors H through J only use their wireless communication channel, these sensors are the most vulnerable to attacks and thus have a low key exchange trust value. Table~\ref{table:ke} list every sensor's key exchange with all sensors in the network of Figure~\ref{fig:network}, e.g., sensor A has a KLJN key exchange with sensors B and D, thus we denote this in set notation as $\mathrm{A}_{\mathrm{kljn}}=\{\mathrm{B},\mathrm{D}\}$. Similarly, sensor A has a wireless key exchange with sensors C, E, F, G, H, I, and J, we denote this in as A$_\mathrm{wireless}=\{\mathrm{C},\mathrm{E},\mathrm{F},\mathrm{G},\mathrm{H},\mathrm{I},\mathrm{J}\}$. Note that $\mathrm{A}_{\mathrm{kljn}} \cap \mathrm{A}_{\mathrm{wireless}} = \emptyset$, that is every sensor communicating with sensor A must be classified as having either a wired KLJN key exchange or a wireless key exchange, but not both. The G key exchange trust system is discussed and analyzed in the following section.

\begin{table}[ht]
\caption{This table list every sensor's key exchange with all sensors in the network of Figure~\ref{fig:network}. Every sensor is classified as having either a wired KLJN key exchange or a wireless key exchange. Set notation is used to categorize the sets as either KLJN or wireless key exchange. }
  \begin{center}
    \begin{tabular}{| c | c | c |}
    \hline
    Sensor & Wired KLJN Key Exchange & Wireless Key Exchange  \\
    \hline
     A & A$_\mathrm{kljn}=\{\mathrm{B},\mathrm{D}\}$ & A$_\mathrm{wireless}=\{\mathrm{C},\mathrm{E},\mathrm{F},\mathrm{G},\mathrm{H},\mathrm{I},\mathrm{J}\}$ \\
     \hline
     B & B$_\mathrm{kljn}=\{\mathrm{A},\mathrm{E}\}$ & B$_\mathrm{wireless}=\{\mathrm{C},\mathrm{D},\mathrm{F},\mathrm{G},\mathrm{H},\mathrm{I},\mathrm{J}\}$ \\
     \hline
     C & C$_\mathrm{kljn}=\{\mathrm{D}\}$  & C$_\mathrm{wireless}=\{\mathrm{A},\mathrm{B},\mathrm{E},\mathrm{F},\mathrm{G},\mathrm{H},\mathrm{I},\mathrm{J}\}$ \\
     \hline
     D & D$_\mathrm{kljn}=\{\mathrm{A},\mathrm{C}, \mathrm{E}\}$ & D$_\mathrm{wireless}=\{\mathrm{B},\mathrm{F},\mathrm{G},\mathrm{H},\mathrm{I},\mathrm{J}\}$ \\
     \hline
     E & E$_\mathrm{kljn}=\{\mathrm{B},\mathrm{D}\}$ & E$_\mathrm{wireless}=\{\mathrm{A},\mathrm{C},\mathrm{F},\mathrm{G},\mathrm{H},\mathrm{I},\mathrm{J}\}$ \\
     \hline
     F & F$_\mathrm{kljn}=\{\mathrm{G}\}$ & F$_\mathrm{wireless}=\{\mathrm{A},\mathrm{B},\mathrm{C},\mathrm{D},\mathrm{E},\mathrm{H},\mathrm{I},\mathrm{J}\}$ \\
     \hline
     G & G$_\mathrm{kljn}=\{\mathrm{F}\}$ & G$_\mathrm{wireless}=\{\mathrm{A},\mathrm{B},\mathrm{C},\mathrm{D},\mathrm{E},\mathrm{H},\mathrm{I},\mathrm{J}\}$ \\
     \hline
     H & H$_\mathrm{kljn}=\emptyset$ & H$_\mathrm{wireless}=\{\mathrm{A},\mathrm{B},\mathrm{C},\mathrm{D},\mathrm{E},\mathrm{F},\mathrm{G},\mathrm{I},\mathrm{J}\}$ \\
     \hline
     I & I$_\mathrm{kljn}=\emptyset$ & I$_\mathrm{wireless}=\{\mathrm{A},\mathrm{B},\mathrm{C},\mathrm{D},\mathrm{E},\mathrm{F},\mathrm{G},\mathrm{H},\mathrm{J}\}$ \\
     \hline
     J & J$_\mathrm{kljn}=\emptyset$ & J$_\mathrm{wireless}=\{\mathrm{A},\mathrm{B},\mathrm{C},\mathrm{D},\mathrm{E},\mathrm{F},\mathrm{G},\mathrm{H},\mathrm{I}\}$ \\
    \hline
    \end{tabular}
  \end{center}
  \label{table:ke}
\end{table}

\subsection{Protocol}

Before sensors can process a KLJN key exchange the KLJN communicators must be authenticated. The authentication of two KLJN units must be completed before they are separated. The KLJN communicator units have a direct wired connection with each other, thus there is no need for networking protocols, only the KLJN key exchange protocol. However, due to the required pre-authentication of the KLJN communicator units the sensor network's topography must be planned ahead.

Since the wired KLJN key exchange has been pre-planned, only the wireless key exchanges need to be processed. Once all sensors in the network has a key exchange with every other sensor in the network, every sensor in the network will classify its key exchange with every peer as being either wired or wireless, e.g., A$_{\mathrm{kljn}}$ and A$_{\mathrm{wireless}}$, B$_{\mathrm{kljn}}$ and B$_{\mathrm{wireless}}$, etc. 

\section{Geometric Key Exchange Trust System}

\subsection{The key exchange trust function}

The geometric key exchange trust system was designed to have a trust function G$_{ij}$ with a range of values, G$_{ij} \in [0,1]$, as a measure of the key exchange trust of sensor $i$ for its communication channel with sensor $j$. The function G$_{ij}$ is for sensor $i$ to evaluate the key exchange trust value of sensor $j$. The input parameters of the function G$_{ij}$ is $i_{\mathrm{kljn}}$, $i_{\mathrm{wireless}}$, $j_{\mathrm{kljn}}$, and $j_{\mathrm{wireless}}$, these parameters are provided by the operator or the base station. 

\subsection{The kill switch}

The kill switch $\gamma_j$ is a binary parameter of sensor $j$ in the G$_{ij}$ function, which is set by the operator to $\gamma_j=0$ when the security of sensor $j$ is compromised, and to $\gamma_j=1$ otherwise. The construction of the G$_{ij}$ function (see below) guarantees that for $\gamma_j=0$ then G$_{ij}=0$.

\subsection{Construction of the key exchange trust function}

When constructing the G$_{ij}$ function, the following goals should be satisfied:

\begin{romanlist}[(ii)]
 \item The contributing terms to the G$_{ij}$ function are determined by:
 \begin{romanlist}[(b)]
  \item The number $K_{ij}$ of mutual KLJN key exchanges with sensors $i$ and $j$, or $K_{ij} = |i_{\mathrm{kljn}} \cap j_{\mathrm{kljn}}|$;
  \item The number $W_{j}$ of KLJN key exchanges with sensor $j$ reduced by $K_{ij}$,  or \\ $W_j=|j_{\mathrm{kljn}} \setminus (i_{\mathrm{kljn}} \cap j_{\mathrm{kljn}})|$;
  \item The number $Z_{j}$ of only wireless key exchanges with sensor $j$ reduced by one (due to the wireless key exchange with sensor $i$), or $Z_j=|j_{\mathrm{wireless}} \setminus i|$.
 \end{romanlist}
 \item Strictly monotonic function. The function G$_{ij}$ is a strictly monotonically increasing function determined by the values $K_{ij}$, $W_j$, $Z_j$. For example, if sensors $j$ and $k$ have values $K_{ij}=K_{ik}$, $W_j=W_k$, and $Z_j > Z_k$, then $\mathrm{G}_{ij} > \mathrm{G}_{ik}$. As a consequence, among the non-compromised sensors, the sensor with a single wireless key exchange should have the lowest contribution.
 
 \item Ranks versus class of connections. The contribution of the term containing $Z_j$ will never exceed the contribution of the term containing $W_j$; and the joint contribution of the terms containing $W_j$ and $Z_j$ will never exceed the contribution of the term containing $K_{ij}$. The reason for this requirement is so that KLJN is the only unconditionally secure key exchange type in the network, thus the rank of its trust is higher. 
\end{romanlist}

Eq.~(\ref{equ:gk3}) utilizes the sum of three geometric series, and satisfies the above conditions:

\begin{equation}
\label{equ:gk3}
\begin{split}
\mathrm{G}_{ij}(\gamma_j) =&
 \begin{cases}
                                      \gamma_j & \text{if } j \in i_{\mathrm{kljn}} \\
                                      \gamma_j \cdot \big(\sum^K_{n=1}(0.3820)^n+\sum^W_{n=1}(0.1729)^n+\sum^Z_{n=1}(0.1474)^n\big) & \text{if }  j \notin i_{\mathrm{kljn}}\\ 
\end{cases}
\end{split}
\end{equation}

\noindent
$\text{with } K = |i_{\mathrm{kljn}} \cap j_{\mathrm{kljn}}| \text{, } W=|j_{\mathrm{kljn}} \setminus (i_{\mathrm{kljn}} \cap j_{\mathrm{kljn}})| \text{, } Z=|j_{\mathrm{wireless}} \setminus i| \text{, and } \gamma_j=\{0,1\}$. The case $\gamma_j=0$ sets G$_{ij}=0$. To satisfy conditions G$_{ij} \leq 1$ and (i) through (iii) above we used the following requirements:

\begin{romanlist}[(ii)]
 \item The third geometric series will saturate at the geometric coefficient of the second series. That is, the third series, in the case of $Z_j \rightarrow \infty$ yields $0.1729$.
 
 \item The sum of the second and third series, will saturate at the geometric coefficient of the first series. That is, in the case of $Z_j \rightarrow \infty$ and $W_j \rightarrow \infty$, their component sum yields $0.3820$. 
 
 \item The sum of the three geometric series will saturate to one. That is, in the case of $Z_j \rightarrow \infty$, $W_j \rightarrow \infty$, and $K_j \rightarrow \infty$, their component sum yields to one. 
\end{romanlist}

\noindent The details of the derivation are shown in the Appendix.

\subsection{Example}

Eq.~(\ref{equ:gk3}) was applied to the network in Figure~\ref{fig:network}. The G key exchange trust values for all the sensors in Figure~\ref{fig:network} are in table~\ref{table:network}. From table~\ref{table:network} some properties of G can be observed. The G function is asymmetric, e.g., in table~\ref{table:network} note that $\mathrm{G}_{BC}(1) \neq \mathrm{G}_{CB}(1)$. There is also incomplete transitive, e.g., in table~\ref{table:network} note that $\mathrm{G}_{AD}(1) = 1$, and $\mathrm{G}_{DC}(1) = 1$, but $\mathrm{G}_{AC}(1) = 0.555$ and does not equal one. Note that the G function given by Eq.~(\ref{equ:gk3}) is unique for the given conditions. The conditions are to have a range between zero and one, and a kill switch. Also note that an infinite number of sensors in lower levels will not undermine a single sensor in a higher level.

\begin{table}[ht]
\caption{This table list $\mathrm{G}_{ij}(\gamma_j)$ key exchange trust values of all the sensors in Figure~\ref{fig:network}. This table assumes $\gamma_j=1$ for all $j$s.}
  \begin{center}
    \begin{tabular}{ c  c | c | c | c | c | c | c | c | c | c | c |}
    \cline{3-12}
    & & \multicolumn{10}{ c| }{$j$} \\
    \cline{3-12}
    &Sensor& A & B & C & D & E & F & G & H & I & J \\
    \cline{1-12}
    \multicolumn{1}{ |c }{\multirow{10}{*}{$i$}} &
    \multicolumn{1}{ |c| }{A} & 1 & 1 & 0.555 & 1 & 0.701 & 0.346 & 0.346 & 0.173 & 0.173 & 0.173 \\
    \cline{2-12}
    \multicolumn{1}{ |c }{} &
    \multicolumn{1}{ |c| }{B} & 1 & 1 & 0.346 & 0.874 & 1 & 0.346 & 0.346 & 0.173 & 0.173 & 0.173 \\
    \cline{2-12}
    \multicolumn{1}{ |c }{} &
    \multicolumn{1}{ |c| }{C} & 0.728 & 0.376 & 1 & 1 & 0.728 & 0.346 & 0.346  & 0.173 & 0.173 & 0.173 \\
    \cline{2-12}
    \multicolumn{1}{ |c }{} &
    \multicolumn{1}{ |c| }{D} & 1 & 0.701 & 1 & 1 & 1 & 0.346 & 0.346 & 0.173 & 0.173 & 0.173 \\
    \cline{2-12}
    \multicolumn{1}{ |c }{} &
    \multicolumn{1}{ |c| }{E} & 0.701 & 1 & 0.555 & 1 & 1 & 0.346 & 0.346 & 0.173 & 0.173 & 0.173 \\
    \cline{2-12}
    \multicolumn{1}{ |c }{} &
    \multicolumn{1}{ |c| }{F} & 0.376 & 0.376 & 0.346 & 0.381 & 0.376 & 1 & 1 & 0.173 & 0.173 & 0.173 \\
    \cline{2-12}
    \multicolumn{1}{ |c }{} &
    \multicolumn{1}{ |c| }{G} & 0.376 & 0.376 & 0.346 & 0.381 & 0.376 & 1 & 1 & 0.173 & 0.173 & 0.173 \\
    \cline{2-12}
    \multicolumn{1}{ |c }{} &
    \multicolumn{1}{ |c| }{H} & 0.376 & 0.376 & 0.346 & 0.381 & 0.376 & 0.346 & 0.346 & 1 & 0.173 &  0.173 \\
    \cline{2-12}
    \multicolumn{1}{ |c }{} &
    \multicolumn{1}{ |c| }{I} & 0.376 & 0.376 & 0.346 & 0.381 & 0.376 & 0.346 & 0.346 & 0.173 & 1 & 0.173 \\
    \cline{2-12}
    \multicolumn{1}{ |c }{} &
    \multicolumn{1}{ |c| }{J} & 0.376 & 0.376 & 0.346 & 0.381 & 0.376 & 0.346 & 0.3458 & 0.173 & 0.173 & 1 \\
   \hline
    \end{tabular}
  \end{center}
  \label{table:network}
\end{table}

As shown in table~\ref{table:network} the G key exchange trust system will give a higher key exchange trust evaluation to sensors that are part of a KLJN key exchange, the more KLJN key exchanges a sensor has the higher the key exchange trust evaluation. Sensors without a KLJN key exchange will have a lower key exchange trust evaluation, even if there are an infinite number of sensors with only wireless key exchange. This mechanism will prevent a lower level sensor attempting to undermine a higher level sensor since there are ceiling limits to sensors that only share a wireless key exchange. A kill switch is in place to allow the G system to remain subjective with any sensor at any time.

\section{Open Questions and Future Work}

Since all sensors in the G system must have both wired and wireless communication channels it will not be practical in some applications. Sensors in the G system will also need to utilize both symmetric encryption for the KLJN key exchange, and asymmetric encryption for the wireless key exchange, this will increase energy requirements, computing requirements, memory, and data storage. Sensors are dependent on the operator or base station to provide or broadcast the KLJN and wireless key exchange sets of every senor in the network, this dependency will require the sensors to remain centralized. For sensors to be autonomous future work must be done where each sensor can broadcast its key exchange sets. Another concern is concealing the cable between the wired sensors. Unconditionally secure key exchange has not been experimented with in sensor networks, but the realization of such a network should be of significant interest. The cost of having unconditionally secure key exchange for sensor networks is high, but such is the price for high security. 

For sensors that cannot communicate with other sensors or the base station due to the distance between them, a multi-hop method is utilized \cite{random1}. The G system does not consider multi-hop cases and would give the sensor a key exchange trust evaluation of the last sensor it was able to communicate with, this can be improved in future work. Sensor networks can also utilize different protocols for different KLJN geometric networks to reduce the cable, time, and KLJN communicators cost as has been analyzed in \cite{me1, me2}.

\section{Conclusion}

In this study we introduced sensor networks along with its applications, limitations, and security issues. We then discuss unconditionally secure key exchanges, and mention how the KLJN key exchange can be included in sensor networks. We also mention current trust methodologies for sensor networks. Since current trust methodologies do not consider unconditionally secure key exchange we introduce the geometric key exchange trust system, a new key exchange trust method for sensor networks that considers unconditionally secure key exchange in the key exchange trust measure. An example of sensor networks with sensors utilizing both wired and wireless communication channels is depicted in Figure~\ref{fig:network}. The G key exchange trust system is then introduced and applied to the sensor network example in Figure~\ref{fig:network}. The G key exchange trust system is then analyzed, discussed, and modeled by Eq.~(\ref{equ:gk3}). Table~\ref{table:network} shows that a higher key exchange trust evaluation is given to sensors with KLJN key exchanges, the more KLJN key exchanges a sensor has, the higher the key exchange trust evaluation. Eq.~(\ref{equ:gk3}) and table~\ref{table:network} also show that there are ceiling limits to sensors that only share a wireless key exchange. The G system depends on the operator or base station to provide the key exchange sets of every sensor in the network. The kill switch allows the G system to remain subjective of every sensor in the network. We then discuss open questions about the G system and possible future improvements.

\appendix

\section{Derivation of G}
 
 The G key exchange trust system has a range from zero to one, and a kill switch. It must also consider an infinite number of sensors, and that a sensor in a lower level cannot undermine a sensor in a higher level. To achieve this we propose to utilized the geometric series since the geometric series can add an infinite sum (or the number of sensors), and equal to a finite value (or one.) Since the highest possible value is one, and with an infinite number of sensors, then the G key exchange trust system of sensor $j$ relative to sensor $i$ can be written as;
\eject

\noindent
 \begin{align}
  \label{a1}
  \mathrm{G}_{ij}(\gamma_j)=\gamma_j \cdot \sum^\infty_{n=1}a^n+b^n+c^n =\gamma_j \cdot 1,
 \end{align}
 

\noindent
with $\gamma_j \in \{0,1\}$ being the kill switch of sensor $j$, and $a$, $b$, and $c$ being the component coefficients. To solve for components $a$, $b$, and $c$, in Eq.~(\ref{a1}) we note that;

\begin{align}
\label{a2}
 \sum^\infty_{n=1}a^n+b^n+c^n =1.
  \end{align}
 

\noindent
Note that the following properties must apply according to the G key exchange trust system. The first property is;

\begin{align}
\label{a3}
 \sum^\infty_{n=1}c^n = b,
  \end{align}
 

\noindent
which means that an infinite number of sensors in the third series ($\sum^\infty_{n=1}c^n$) cannot undermine a single sensor in the second series ($b=\sum^1_{n=1}b^n$.) The second property is;

\begin{align}
\label{a4}
 \sum^\infty_{n=1}b^n + c^n = a,
  \end{align}
 

\noindent
which means that an infinite number of sensors in the second series ($\sum^\infty_{n=1}b^n$,) and an infinite number of sensors in the third series ($\sum^\infty_{n=1}c^n$) cannot undermine a single sensor in the first series ($a=\sum^1_{n=1}a^n$.) Eq.~(\ref{a3}) and Eq.~(\ref{a4}) can be rewritten to isolate the infinite summation of $b$ as follows,

\begin{align*}
\sum^\infty_{n=1}b^n = a - b.
\end{align*}


\noindent
Also, note that if $r \in \mathbb{R} : |r|<1 $ then $\sum^\infty_{n=1}r^n = \frac{r}{1-r}$. Given these properties Eq.~(\ref{a2}) can be derived as;

\begin{align*}
 \sum^\infty_{n=1}a^n+b^n+c^n =1, \\
 \sum^\infty_{n=1}a^n+\sum^\infty_{n=1}b^n+\sum^\infty_{n=1}c^n =1, \\
 \sum^\infty_{n=1}a^n+\big(a-b\big)+\big(b\big)=1, \\
 \sum^\infty_{n=1}a^n+a=1, \\
 \frac{a}{1-a}+a=1. \\
\end{align*}

\noindent
The resulting equation $a/(1-a)+a =1$, can be solved for $a$ by using the quadratic formula giving values $a=(3-\sqrt{5})/2$ and $a=(3+\sqrt{5})/2$. Since $|a|<1$, then the only converging value is $a=(3-\sqrt{5})/2$. Thus the component $a$ is,

\begin{equation}
a=\frac{3-\sqrt{5}}{2} \approx 0.3820. 
\label{a5}
\end{equation}

\noindent
A similar method can be used to solve for $b$ and $c$ in Eq.~(\ref{a2}). 

To solve for $b$ note that 

\begin{equation*}
 \sum^\infty_{n=1}b^n =a-b, \\
\end{equation*}
\begin{equation}
 \frac{b}{1-b}=a-b. \\
 \label{a6}
\end{equation}

\noindent
Solving for $b$ in Eq.~(\ref{a6}) gives two solutions. The converging solution is,

\begin{equation}
 b=\frac{a+2+\sqrt{a^2+4}}{2}.
 \label{a7}
\end{equation}

\noindent
Given Eq.~(\ref{a5}) and substituting for $a$ in Eq.~(\ref{a7}) gives,

\begin{equation}
 b=\frac{7-\sqrt{5}-\sqrt{30-6\sqrt{5}}}{4} \approx 0.1729.
 \label{a8}
\end{equation}

\noindent
Thus the component $b$ is given by Eq.~(\ref{a8}).

The component $c$ can be solved by utilizing Eq.~(\ref{a3}). Note that,

\begin{equation*}
  \sum^\infty_{n=1}c^n = b,\\ 
\end{equation*}
\begin{equation}
 \frac{c}{1-c}=b. \\
 \label{a9}
\end{equation}

\noindent
Given Eq.~(\ref{a8}) and substituting for $b$ in Eq.~(\ref{a9}), then solving for $c$ will give,

\begin{equation}
 c=\frac{\sqrt{30-6\sqrt{5}}+\sqrt{5}-7}{\sqrt{30-6\sqrt{5}}+\sqrt{5}-11} \approx 0.1474.
 \label{a10}
\end{equation}

\noindent
Thus the component $c$ is given by Eq.~(\ref{a10}).

The derivations above were derived to consider any number of sensors, thus the G key exchange trust function holds for zero sensors to an infinite number of sensors. In reality there will be a limited number of sensors in a network. 

The component $a$ will only consider sensors that are conditionally secured with mutual KLJN key exchanges, e.g., if sensor $i$ and sensor $j$ have mutual KLJN key exchanges with third parties, then this can be written in set notation as the intersection of sensor $i$'s $i_{\mathrm{kljn}}$ set and sensor $j$'s $j_{\mathrm{kljn}}$ set. This can be expressed as $i_{\mathrm{kljn}} \cap j_{\mathrm{kljn}}$. The number of mutual KLJN key exchanges with third parties between sensors $i$ and $j$ can be expressed as $K=|i_{\mathrm{kljn}} \cap j_{\mathrm{kljn}}|$. Thus, there are $K$ mutual sensors between sensors $i$ and $j$. 

The component $b$ will only consider sensors that are conditionally secured without mutual KLJN key exchanges, e.g., if sensor $i$ evaluates the number of key exchanges in sensor $j$, then only the number of KLJN key exchanges in sensor $j$ that do not have mutual KLJN key exchanges with sensor $i$ will be noted. This can be expressed as $W=|j_{\mathrm{kljn}} \setminus (i_{\mathrm{kljn}} \cap j_{\mathrm{kljn}})|$. The purpose of having component b is based on the belief that a sensor with a KLJN key exchange should have a higher key exchange trust value than a sensor without a KLJN key exchange.

The component $c$ will only consider sensors that are conditionally secured with only wireless key exchanges, e.g., if sensor $j$ only has wireless key exchanges with other sensors then the number of sensors that can verify a wireless key exchange with sensor $j$ is $Z=|j_{\mathrm{wireless}} \setminus i|$. 

The G key exchange trust system can evaluate the key exchange trust level of sensor $j$ relative to sensor $i$, this can be expressed as G$_{ij}(\gamma_j)$, with $\gamma_j$ being the kill switch for sensor $j$. $\mathrm{G}_{ij}(\gamma_j)$ can be expressed as the following equation;

\begin{equation*}
\begin{split}
\mathrm{G}_{ij}(\gamma_j) =&
 \begin{cases}
                                      \gamma_j & \text{if } j \in i_{\mathrm{kljn}} \\
                                      \gamma_j \cdot \big(\sum^K_{n=1}(0.3820)^n+\sum^W_{n=1}(0.1729)^n+\sum^Z_{n=1}(0.1474)^n\big) & \text{if }  j \notin i_{\mathrm{kljn}}\\ 
\end{cases}
\end{split}
\end{equation*}

\noindent
$\text{with } K = |i_{\mathrm{kljn}} \cap j_{\mathrm{kljn}}| \text{, } W=|j_{\mathrm{kljn}} \setminus (i_{\mathrm{kljn}} \cap j_{\mathrm{kljn}})| \text{, } Z=|j_{\mathrm{wireless}} \setminus i| \text{, and } \gamma_j=\{0,1\}$.

\end{document}